\documentclass[11pt,a4paper]{article}

\usepackage{graphicx}
\usepackage{url}
\usepackage{authblk}
\usepackage{siunitx}
\usepackage [utf8]{inputenc}
\author[1]{Alejandro Laso Garcia \thanks{Corresponding author: a.garcia@hzdr.de}}
\author[2]{Mikhail Mishchenko}
\author[1]{Victorien Bouffetier}
\author[3]{Gabriel Pérez-Callejo}
\author[2]{Karen Appel}
\author[4]{Alexey Arefiev}
\author[1]{Carsten Baehtz}
\author[2]{Erik Brambrink}
\author[5]{Mihail Cernaianu}
\author[5]{Domenico Doria}
\author[1,14]{Tobias Dornheim}
\author[6]{Gillis M. Dyer}
\author[7]{Nicolas Fefeu}
\author[6]{Eric Galtier}
\author[14,1]{Thomas Gawne}
\author[5]{Petru V. Ghenuche}
\author[2]{Sebastian Goede}
\author[8]{Johannes Hagemann}
\author[1,9]{Marie-Luise Herbert}
\author[1]{Hauke Höppner}
\author[1]{Lingen Huang}
\author[2]{Oliver Humphries}
\author[12]{Mae Jones}
\author[6]{Dimitri Khaghani}
\author[1]{Thomas Kluge}
\author[2]{Jayanath Koliyadu}
\author[1,9]{Dominik Kraus}
\author[6]{Hae Ja Lee}
\author[9]{Julian Lütgert}
\author[2]{Mikako Makita}
\author[9]{Jean-Paul Naedler}
\author[6]{Bob Nagler}
\author[2]{Motoaki Nakatsutsumi}
\author[6]{Quynh Nguyen}
\author[1]{Alexander Pelka}
\author[2]{Thomas R. Preston}
\author[9]{Chong Bing Qu}
\author[2]{Sripati V. Rahul}
\author[10]{Lisa Randolph}
\author[9]{Ronald Redmer}
\author[1]{Martin Rehwald}
\author[11]{Hans G. Rinderknecht}
\author[2]{Angel Rodriguez-Fernandez}
\author[7]{Joao J. Santos}
\author[1]{Ulrich Schramm}
\author[1]{Michal Smid}
\author[8]{Cornelius Strohm}
\author[2,13]{Jergus Strucka}
\author[2]{Minxue Tang}
\author[2,8]{Patrik Vagovic}
\author[1]{Milenko Vescovi}
\author[1]{Long Yang}
\author[1]{Karl Zeil}
\author[2]{Ulf Zastrau}
\author[1]{Thomas E. Cowan}
\author[1]{Toma Toncian}

\affil[1]{Helmholtz-Zentrum Dresden - Rossendorf, Dresden, Germany}

\affil[2]{European XFEL, Schenefeld, Germany}

\affil[3]{Departamento de Física Teórica, Atómica y Óptica, Universidad de Valladolid, Valladolid, Spain}

\affil[4]{Department of Mechanical and Aerospace Engineering, University of California at San Diego, La Jolla, USA}

\affil[5]{Extreme Light Infrastructure - Nuclear Physics, IFIN-HH, Măgurele, Romania}

\affil[6]{SLAC National Accelerator Laboratory, Menlo Park, USA}

\affil[7]{CNRS - Université de Bordeaux, CELIA, Talence, France}

\affil[8]{DESY Deutsches Elektronen-Synchrotron, Photon Science, Hamburg, Germany}

\affil[9]{Universität Rostock, Institut für Physik, Rostock, Germany}

\affil[10]{Universität Siegen, Department Physik, Siegen, Germany}

\affil[11]{Laboratory for Laser Energetics, University of Rochester, Rochester, USA}

\affil[12]{The University of Edinburgh, School of Physics and Astronomy, Edinburgh, United Kingdom}

\affil[13]{Plasma Physics Group, Imperial College London, London, United Kingdom}

\affil[14]{Center for Advanced Systems Understanding (CASUS), G\"orlitz, Germany}

\begin{document}


\title{XFEL Imaging techniques for High Energy Density and Inertial Fusion Energy Research at HED-HiBEF}
\maketitle


\begin{abstract}
The imaging platform developed at the High Energy Density - Helmholtz International Beamline for Extreme Fields (HED-HiBEF) instrument at the European XFEL and its applications to high energy density and fusion related research are presented. The platform combines the XFEL beam with the high-intensity short-pulse laser ReLaX and the high-energy nanosecond-pulse laser DiPOLE-100X. The spatial resolution is better than 500 nm and the temporal resolution of the order of 50 fs. We show examples of blast waves and converging cylindrical shocks in aluminium, resonant absorption measurements of specific charged states in copper with ReLaX and planar shocks in polystyrene material generated by DiPOLE-100X. We also discuss the possibilities introduced by combining this imaging platform with a kJ-class laser.

\end{abstract}

\section{Introduction}

Probing high energy density states of matter is a challenging task. The short time scales (from fs to tens of ns), the small spatial scales (of nanometer to tens of micrometers) and the high electron densities require the use of short, bright x-ray beams. Several variations of x-ray imaging have been developed depending on the sample parameters (size, optical thickness, index of refraction for the used x-ray energy): absorption radiography is used for optically thick targets where absorption is dominant, phase-contrast imaging for optically thin samples where absorption is negligible, grating interferometry and Talbot grating interferometry to further increase sensitivity and accuracy by exploiting the Talbot effect.

In optical laser facilities, x-ray imaging has been mostly implemented via x-ray backlighters \cite{landen_2001, barrios_2014, turk_2010, marshall_2021, bouffetier_2024}, providing novel insight on the dynamic of shock generation and propagation in planar geometries at LULI  \cite{antonelli_2017, rigon_luli_2021}),  in direct drive configuration at OMEGA  \cite{antonelli_2024} and at the National Ignition Facility, imaging of capsule implosions has even provided crucial information on the influence of the tent holder \cite{tommasini_2015}. Two-grating interferometry has also been developed, with a phase grating and an analyzer grating \cite{wegert_2024}.  However, all these techniques have limited temporal resolution, down to the picosecond range, due to the duration of the laser generated x-ray pulse, or the gating duration on x-ray framing cameras \cite{do_2021}. They also suffer from limitations on the spatial resolution, typically in the range of micrometers. Imaging with the betatron emission from electrons in a laser-wakefield accelerator has also been demonstrated  \cite{wood_2018}, and show enhanced spacial and temporal resolution, with a limited photon flux compared to hard X-ray FELs.

The advent of hard X-ray Free Electron Lasers (XFELs) has overcome these limitations. The x-ray beam generated in an XFEL has a high spatial and temporal coherence, a short pulse duration ($<$\,50 fs), narrow bandwidth (eV) and an extreme peak brightness ($10^{33}$ photons s$^{-1}$ mrad$^{-2}$ mm$^{-2}$ / 0.1 \%BW \cite{tschentscher_2017}). Using these high-quality beams to image high energy density states has pushed the boundaries of spatial and temporal resolution, leading to discovery of never-seen-before phenomena that would not have been resolvable with optical backlighters otherwise. At the Matter at Extreme Conditions station at LCLS, the x-ray imaging platform \cite{nagler_2016, galtier_2025} has been used to probe shocks in diamond \cite{schropp_2015} and low density foams \cite{parisuana_2025}, the interplay of void in materials and shock propagation \cite{hodge_2025} and multi-frame imaging \cite{hodge_2022}. At SACLA, imaging has been used to elucidate electron transport dynamics in solid foils and nanowire arrays \cite{tanaka_2025}, shock splitting in diamond \cite{makarov_2023} and to elucidate turbulent spectra from Rayleigh-Taylor instabilities with unprecedented resolution \cite{rigon_2021}. At EuXFEL, a new pathway to achieve high-pressure states via convergent shockwaves was found in micrometer-sized wires \cite{lasogarcia_2024}. 

In this paper, we describe the experimental imaging platform developed at the High Energy Density - Helmholtz International Beamline for Extreme Fields (HED-HiBEF) instrument at EuXFEL \cite{zastrau_2021}. We will discuss the technical aspects and the resolution limitations. We will show examples of physical processes generated with the high-intensity short-pulse laser, ReLaX \cite{lasogarcia_2021}, and the high-energy nanosecond-pulse laser DiPOLE-100X \cite{Mason}. Finally, we examine the prospects of using this platform for fusion relevant research with the current capabilities at HED-HiBEF and when coupling it to a kJ-class laser.

\section{The imaging platform at HED-HiBEF}
The standard hard x-ray imaging setup at the HED-HiBEF instrument consists of a set of beryllium compound refractive lenses (CRLs) and a high-magnification optical microscope with a scintillator screen. The CRL sets are adapted to the x-ray energy as well as the desired magnification. The typical parameters of the CRLs are stacks of 15 to 34 lenses (each lens has a curvature radius of \SI{50}{\um} and an aperture of \SI{400}{\um}), with focal ranges covering 15 - 35 cm and magnification of 15-34. Up to now, the energy range covered has been from 7 keV to 8.5 keV; however, this can be extended to higher x-ray energies. The high-magnification detector consists of a scintillator (GAGG, Lu:Ce, YAG) coupled to a CMOS detector (Andor Zyla) via optical objectives with a selectable magnification of 2x, 7.5x and 10x \cite{koliyadu_2025}. The highest x-ray and optical magnification of 340 results in an equivalent pixel size on target down to $18.3 \pm 0.1$ nm based on calibrated target measurements.

\begin{figure}[!h]
\includegraphics[width=\textwidth]{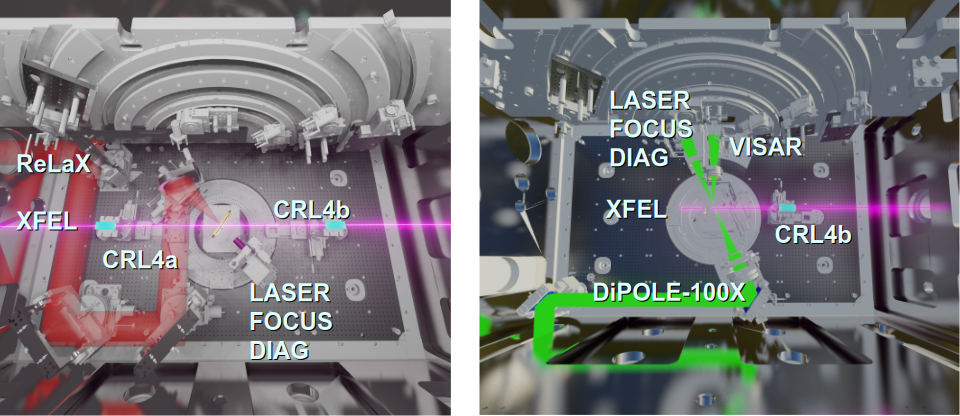}
\caption{X-ray imaging setups: a) in combination with ReLaX and b) in combination with DiPOLE-100X. The detector is located outside the interaction chamber to the right in both images.}
\label{figure:fig_setups}
\end{figure}

The lenses are located at a distance after the target such that an image is formed on the scintillator downstream. The distance from the target to the detector is flexible ranging from 2 m to 6.3 m. The lenses are mounted on a hexapod to allow precise alignment of the stack in 6 degrees of freedom. The hexapod is located on top of two linear stages, one transversal to the beam, for a quick switch of in and out lens position, and one longitudinal to the beam to scan along the x-ray axis. The usual travel range along the x-ray axis is about $\pm$\SI{10}{cm}. This holder is compatible with either ReLaX or DiPOLE-100X beam routings (as shown in Figure \ref{figure:fig_setups}, with the imaging CRL labeled CRL4b), as well as additional diagnostics like x-ray spectroscopy. Additionally, a twin CRL (CLR4a in Figure \ref{figure:fig_setups}) stack can be placed in front of the target to generate a sub-micrometer focus on target, or use a point-projection imaging scheme. Both sets can be used simultaneously to measure the focal spot size, or monitor the overlap in experiments using the two-color XFEL mode. 

The intensity at the detector can be described via the transport of intensity equation \cite{teague_1982}:
\begin{equation}
I(x,y, \Delta z) = I(x,y) - \frac{\Delta z}{k} \nabla \cdot [I(x,y) \nabla \Phi(x,y)]
\end{equation}
with $I(x,y)$ the intensity in the plane at the object exit (what is known as 'contact plane'), $\Delta z$ is the x-ray propagation distance from the target until the detector, $k$ is the x-ray wave number and $\Phi (x,y)$ is the phase of the object. In this equation, the index of refraction $n = 1 - \delta + i \beta$ of the object is encoded: the absorption is related to $\beta$ since $I(x,y) = I_0 e^{(- 2 k \int \beta(x,y,z') dz')}$ and the phase is $\Phi = k \int \delta(x,y,z') dz'$, with the integration through the target thickness, along the x-ray axis. Therefore, it is possible to extract the object index of refraction under certain assumptions using phase-retrieval methods \cite{paganin_2002, paganin_2020, huhn_2022}. However, single-shot reconstruction of the object suffers in laser-plasma experiments from multiple issues: the uncertainty in the illumination function with variation in intensity and jitter, the plasma generated in the interaction and the non-uniformities introduced by the CRLs can affect the convergence of the reconstruction methods.

Most of the problems previously mentioned can be mitigated and the sensitivity of the system enhanced by means of Talbot interferometry. This technique has been successfully employed at laser facilities with x-ray backlighters \cite{valdivia_2018, bouffetier_2020, valdivia_2021, perezcallejo_2023} and other XFELs \cite{bouffetier_thesis_2021, parisuana_2025, galtier_2025}. Talbot interferometry is based on the Talbot effect. To record the interferometric data, a detector is placed in an arbitrarily chosen self image Talbot plane. Disturbance of the interferometric pattern by the introduction of a sample in the x-ray path can thereafter be linked to transmission, differential phase, and dark field radiographs through Fourier analysis \cite{perezcallejo_2022}.

In a more formal way, the intensity at the detector can be expressed as \cite{valdivia_2018, bouffetier_2020}:
\begin{equation}
I(x,y) = A(x,y) + B(x,y) e^{i \phi(x,y)}
\end{equation}
where $A(x,y)$, $B(x,y)$ and $\phi(x,y)$ are real functions that can be extracted and related to the attenuation, the fringe visibility and the phase. A Fourier analysis using an image with the object and an image without the object allows the extraction of the object properties as attenuation $A= A_{obj}/A_{ref}$ and $\phi = \phi_{obj} - \phi_{ref}$.  In this case, the problem has shifted from analyzing the target image itself, to analyzing the variations on the periodicity and amplitude of the grating peaks, without any assumption on the object itself. The grating used in our case is a checkerboard diamond grating with a phase shift of $\pi$/2 at 8 keV, horizontal pitch of \SI{9.2}{\um} and thickness of \SI{10}{\um}.

\begin{figure}[!h]
\includegraphics[width=\textwidth]{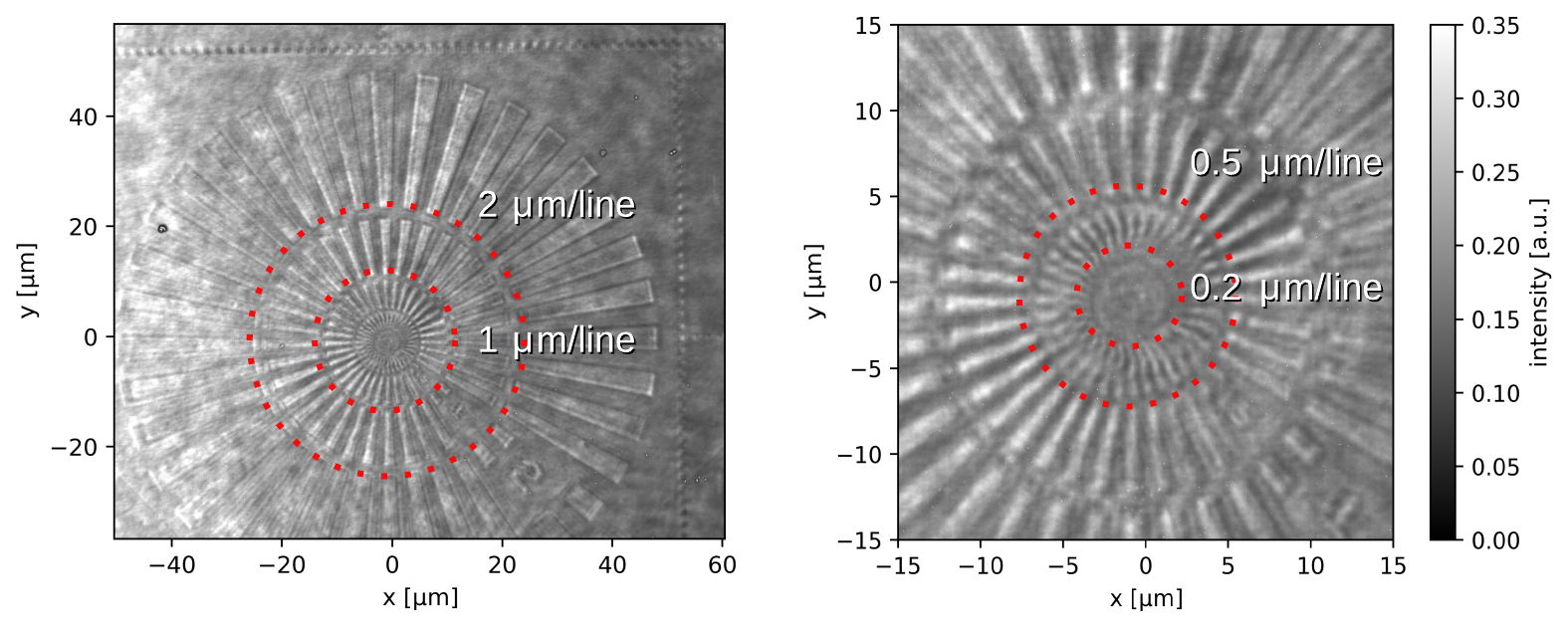}
\caption{Resolution target for x-ray imaging. The circular areas show the dimensions of each individual line at that radius.}
\label{figure:fig_siemens_star}
\end{figure}

The resolution of the imaging system was measured using a calibration target (XRESO-50HC from NTT). It consists of a Siemens Star made of tungsten with a thickness of 500 nm. The minimum spatial features are 50 nm at the centermost past of the patterned area. As seen in Figure \ref{figure:fig_siemens_star}, sub-\SI{500}{nm} structures are resolved for the highest magnification configuration.

\section{Imaging results examples}
Here we showcase a series of examples of the phenomena that have been investigated with these techniques.

\subsection{Blast waves generated with a short-pulse laser}
The high-intensity short-pulse interaction with the target generates a hot spot of the order of a few micrometers. The high temperature gradient between this spot and the surrounding material leads to the generation of a blast wave. Previous studies of isochorically heated solids have characterized the hot electron source as well as the preheating and velocity of the wave \cite{santos_2007, jakubowska_2017, santos_2017, batani_2024}. They have also used simulations to infer the shock pressures in the range of 100 Mbar. The high energy density states generated in this fashion show great promise towards the study of astrophysical phenomena in the laboratory.

\begin{figure}[!h]
\includegraphics[width=\textwidth]{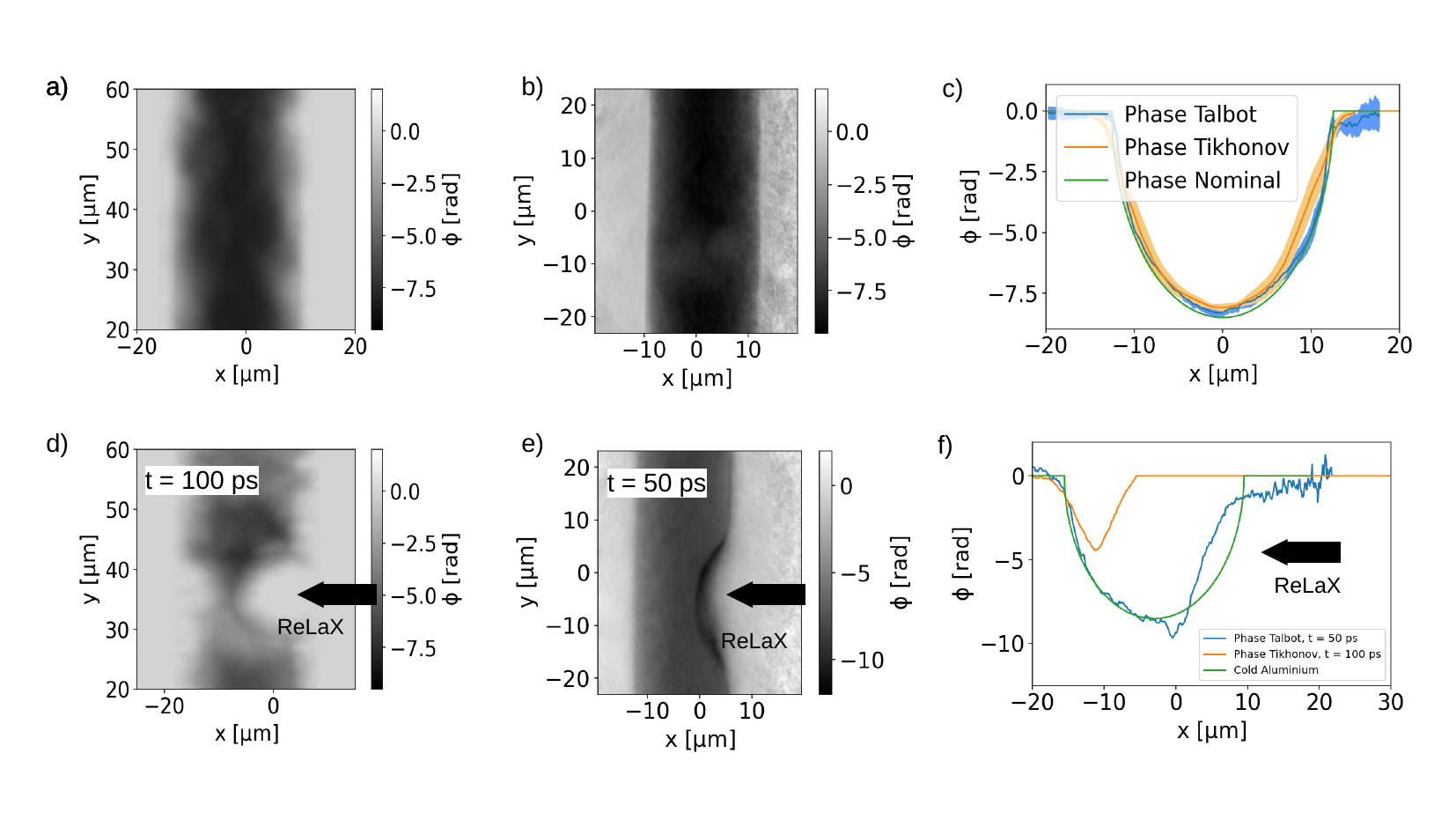}
\caption{Retrieved phase for aluminium wires. For PCI configuration: a) x-ray only imaging of the unpumped wire, d) x-ray imaging 100\,ps after ReLaX arrival. For Talbot imaging configuration: b) x-ray only for cold aluminium wire, e) talbot imaging 50\,ps after ReLaX arrival. Panel c) shows a lineout at the central part of the both PCI and Talbot configuration images and the expected theoretical phase shift from a perfect \SI{25}{\mu} aluminium wire. Panel f) shows a horizontal lineout at the central part of the shock for both cases. In the lower panels the thick black arrow represents the projected incoming direction of ReLaX in the x-ray imaging plane.}
\label{figure:fig_aluminium_sedov}
\end{figure}

The platform was used in combination with ReLaX to study shocks propagating in wire targets. Here, we show a comparison between PCI and Talbot imaging on aluminium wires of 25 $\mu$m. In Figure \ref{figure:fig_aluminium_sedov}, we show both the reconstructed phase for an undriven target, as well as the interaction of ReLaX and the target.

In the case of PCI (displayed in Figure \ref{figure:fig_aluminium_sedov} a), the x-ray magnification was $M_x = 15$ and the detector optical magnification $M_o = 2$, corresponding to a total magnification of 30 and an equivalent pixel size on target of 216 nm/px. We use a non-linear phase-retrieval algorithm in the near-field regime with Tikhonov regularization \cite{huhn_2022} as implemented in the \emph{HoToPy} package for python \cite{lucht_2025}. The constraints used were a single element object (aluminium) with a $\beta/\delta = 1.63 \times 10^{-3}$, a support indicating where the object was located and a non-positive phase. The Fresnel number for this configuration was $Fr = 0.012$. With these settings, the reconstructed phase shift for the aluminium wire at the center is 8.3 rad.

In the case of Talbot imaging, the x-ray magnification was $M_x = 34$ and the detector magnification $M_o = 10$. The total magnification was 340 with an equivalent pixel size on target of 18 nm/px. The phase was retrieved using the TIA/TNT algorithm \cite{perezcallejo_2022}. The data shown here correspond to a shot with a pump-probe delay of 0 ps. Essentially, the areas away from the laser focus still remain at cold temperature and no hydrodynamic motion has taken place yet. A lineout comparison between both phase retrieval methods and with the phase shift expected from a perfect aluminium wire is shown in panel c) showing a good agreement between the methods.

The differences between both imaging techniques become clear when studying blast waves generated by the ReLaX interaction with the target. For the PCI case, the recorded signal around the shock region contains a mixture of the imaging of CRL impurities together with the plasma halo. These features cannot be correctly phase-retrieved. Furthermore, the shock phase-shift obtained is lower than that of cold aluminium. This limitation can be overcome by using Talbot imaging. Using the Talbot grating as an interferometer, effectively provides a filter that compensates spurious effects not related to the object imaging. This can be seen on how the Talbot phase maps are more homogeneous across the whole wire when compared to the PCI phase. A lineout along the central part of the shock is shown in panel f). The correct phase projection is retrieved and an increased phase-shift shows the location of the shock front.

\subsection{Wire compression driven by return currents}
The first implementation of the imaging platform with ReLaX enabled the discovery of a compression wave in micrometer-thin wires \cite{lasogarcia_2024}. Follow-up experiments have demonstrated the robustness of this compression method by using different materials \cite{long_2025}. In this process, the hot electrons expelled from the target generate a charge imbalance that drives an intense return current. This return current lasts in the order of the laser pulse duration and is restricted to a skin depth of less than \SI{1}{\um}. This surface is heated to high temperatures, up to hundreds of eV. The gradient of the surface temperature to the colder inner material leads to an ablation driven cylindrical wave that travels towards the wire axis. At the convergence point of the wave, a 10x compression has been demonstrated with simulations predicting pressures up to 800 Mbar on copper \cite{lasogarcia_2024}.

\begin{figure}[!h]
\includegraphics[width=\textwidth]{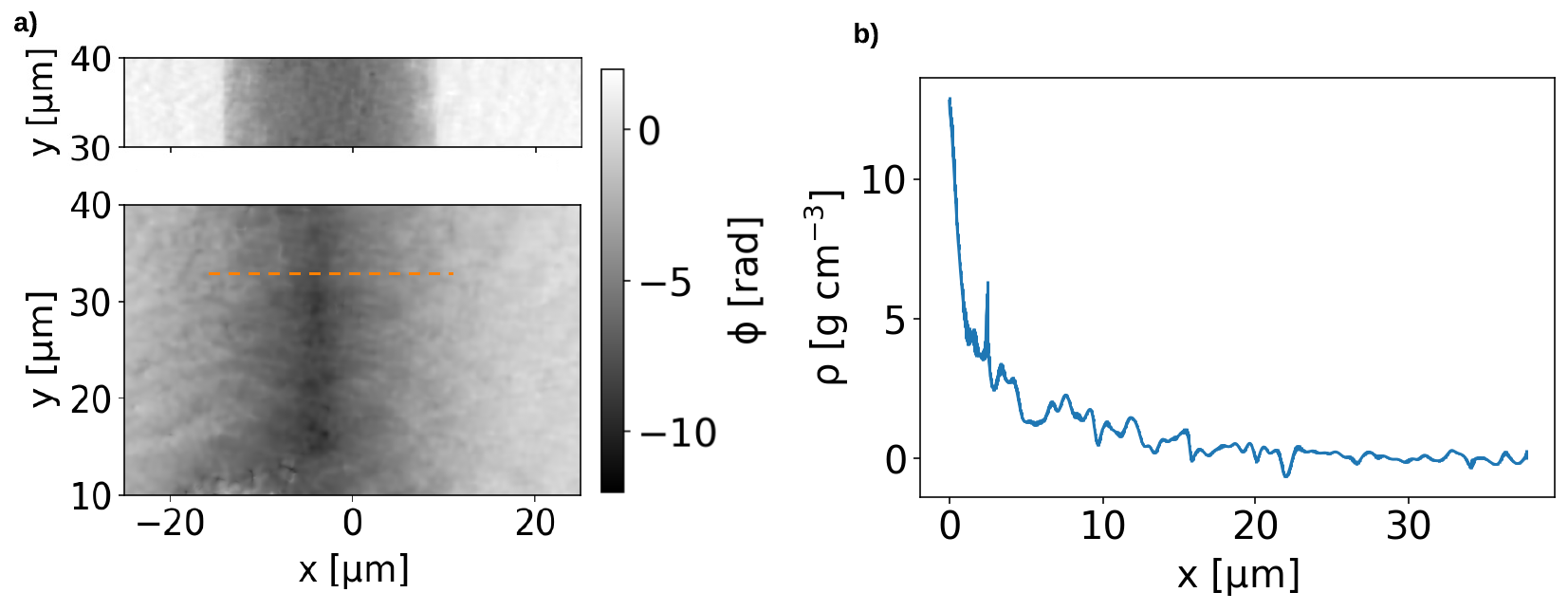}
\caption{a) Phase maps for an aluminium wire, with the upper panel showing the cold target and the lower panel the target 700 ps after ReLaX arrival. The orange dashed line indicates the point along the wire where the compression is maximal. b) Abel inverted density at the highest compression point.}
\label{figure:fig_aluminium_convergence}
\end{figure}

Here we show results of the compression of a \SI{25}{\um} aluminium wire with Talbot imaging. The phase retrieval for a cold wire is displayed in the upper part of Figure \ref{figure:fig_aluminium_convergence} a), while the phase shift for a pump-probe delay of 700 ps is shown in the lower panel. For this delay, the compression wave reached the wire axis at a distance of approximately \SI{45}{\um}. The phase retrieved image was Abel inverted using a deconvolution procedure developed by Daun et al. \cite{daun_2006} and implemented in the \emph{pyabel} package. The density shown on panel b) exhibits a maximum at the wire axis after an abel inversion, corresponding to a density $\rho$ = \SI{12.7}{g\,cm^{-3}}. This calculated compression factor is 4.7. 

\subsection{Charge-state specific resonant imaging}
\label{sec:resonant_imaging}
The narrow energy bandwidth of the XFEL beam enables investigations of energy-selective process. In particular, specific electron transitions between the K and L shells can be probed by using ReLaX to ionize the target and using the XFEL beam to excite the electrons. In our case, copper foils with thickness of \SI{2}{\um} and \SI{5}{\um} were used as targets. We chose an XFEL energy of 8.163 keV, corresponding to the energy transition from the K-shell to the L-shell of Cu$^{21+}$ ions. Previous experiments have performed studies of resonant absorption with spectroscopy and x-ray imaging \cite{huang_2025, huang_arxiv_2025}.

\begin{figure}[!h]
\includegraphics[width=\textwidth]{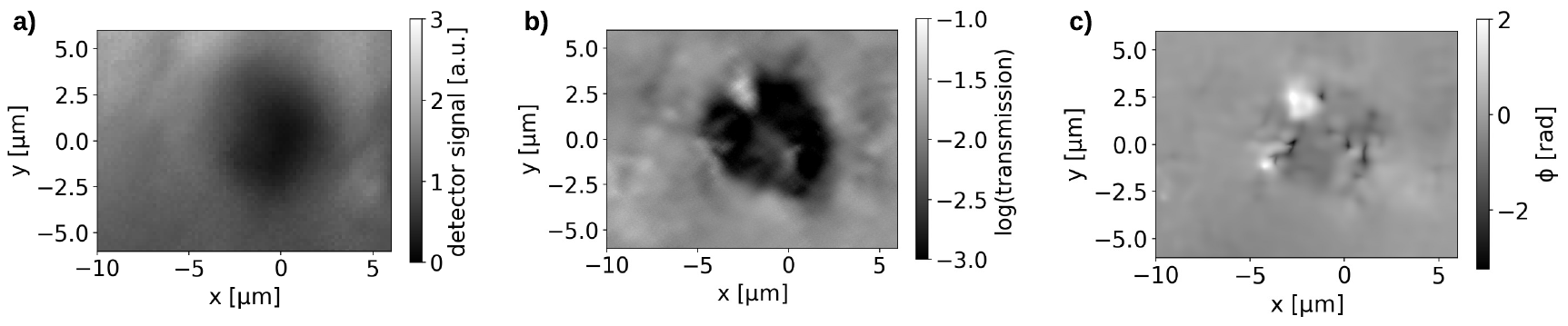}
\caption{a) Flatfielded x-ray image of a \SI{5}{\um} copper foil, 200 fs after ReLaX arrival. b) Attenuation map for a \SI{2}{\um} copper foil 4.8 ps after ReLaX main pulse. c) Phase map for the same \SI{2}{\um} foil and time step.}
\label{figure:fig_copper_resonant}
\end{figure}

The flatfielded image of a \SI{5}{\um} thick copper foil upon ReLaX irradiation can be seen in Figure \ref{figure:fig_copper_resonant} a). The delay between ReLaX and X-ray was 200 fs. A decrease in detector counts at the center of the image, with a width comparable to the laser focus is apparent at the center of the image. However, this attenuation is a mixture of absorption and refraction effects. Deconvolving the actual attenuation would require making an assumption, a priori, on the $\beta / \delta$ ratio for copper. As this is a resonant process, the value of $\beta$ is highly dependent on the ionization state of the copper foil. The ionization state itself is not spatially homogeneous, adding further complications to any phase-retrieval attempt. In the case of Talbot imaging, without assumptions on the target itself, the situation is different. Here we show an example of the transmission map and phase map for a pump-probe delay of 4.8 ps. Figure \ref{figure:fig_copper_resonant} b) and c) show the reconstructed maps. The phase shift displayed corresponds to the phase delay with respect to a cold copper foil, thus a null phase shift is shown across the area with a deviation where the laser impacted on the foil. Similarly, the transmission map shows the transmission with respect to a cold foil. There is an ambiguity due to the XFEL pulse energy used to measure the cold foil and the laser-shot foil. To account for this, we use a relative measurement of the transmission between the laser irradiated area, and an area far outside. This ratio of transmitted x-rays inside the laser area to outside is ~0.3. If the effect were to be volumetric through the complete foil thickness, the mass attenuation coefficient would be 613 cm$^2$ g$^{-1}$. For a cold foil of the same thickness, the mass attenuation coefficient at this x-ray energy is 48 cm$^2$ g$^{-1}$. This is a clear indication that the x-rays are being resonantly absorbed.

\subsection{Shock-compression of polystyrene with DiPOLE-100X}
Shock compression is a standard technique to study high-pressure states and phase transitions. Most shock compression studies employ x-ray diffraction and velocity interferometer system for any reflector (VISAR) techniques to extract the conditions reached during compression. However, in the case of VISAR, a limitation is given in the shock conditions that can be studied due to the reflectivity needed as well as the impossibility of probing shocks during propagation inside optically opaque targets.

Employing imaging to study the propagation of a laser generated shock in a material extends the conditions to be studied, and provides snapshots of the shock development during and after laser irradiation. We used a target made of polystyrene, with an aluminium ablator deposited on one face. The polystyrene block had dimensions $0.475 \times 0.475 \times 5$ mm$^3$, the aluminium ablator was $0.475 \times 0.010 \times 5$ mm$^3$. The DiPOLE laser shot on the ablator side at an angle \ang{22.5} with respect to the laser normal, launching a shock through the aluminium and polystyrene materials. The x-ray probed transversally to the shock propagation. DiPOLE was used at 2$\omega$ (\SI{515}{nm}), with 35 J on target and a phase plate of \SI{300}{\um}.

\begin{figure}[!ht]
\includegraphics[width=\textwidth]{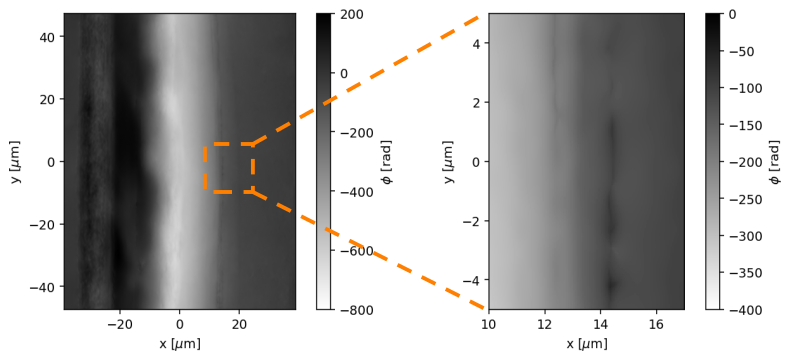}
\caption{Left: Phase shift map for a shocked polystyrene block with respect to an unshocked block. Right: zoom in section around the front shock showing multiple phase sharp gradients.}
\label{figure:fig_dipole_shock}
\end{figure}

The phase retrieved image in Figure ~\ref{figure:fig_dipole_shock} a) shows the richness of the shock process. The phase is calculated with respect to the undriven target. The ablator is seen in the area $x<$ \SI{-20}{\um}. The shock front appears in the range of 12 $< x <$ 16 \SI{}{\um}. There, we observe the presence of at least two distinct discontinuities in the phase. This example shows the enhanced capabilities of imaging shocks against other methods: the possibility to observe the shock front, and possibly rarefaction waves as they propagate inside the target. Recent theoretical studies have analyzed the assumptions and effect of an ablator thickness and impedance mismatch in laser-loading experiments \cite{duchateau_2025}. Such theoretical studies could obtain experimental validation with this imaging platform.

\section{Research potential for IFE applications}
Currently, a demonstration of a target gain $>1$ has only been demonstrated at the NIF \cite{shawareb_2024}. This achievement was the culmination of decades of improvements in laser design, targetry, diagnostics and simulations. An overview of the challenges overcome can be found in \cite{hurricane_2024, hurricane_2025}. 

The conditions needed to reach ignition are extreme, thus an implosion facility needs the most energetic laser available such as NIF and LMJ. However, on the path to ignition and inertial fusion energy, challenges can be tackled with smaller lasers, currently available at XFEL facilities. For example, inhomogeneities in the ablator can lead to degradation in performance \cite{casey_2021}. Studies performed at MEC and HED-HiBEF have demonstrated diamond formation from plastic ablators, similar to those used in NIF capsules, when irradiated with a long pulse laser \cite{kraus_2017, frost_2024}. Hydrodynamic instability control has been, and remains, a challenge. The ability to observe the instability growth holds great promise towards a more effective suppression. Studies of instabilities and turbulent regime have been performed at SACLA \cite{rigon_2021}. New nano-accelerator concepts are also gaining momentum, such as nanowire arrays for direct drive laser fusion. Experiments aimed to understanding nanowire evolution under irradiation have been started \cite{tanaka_2025}.

The inherent value of performing imaging experiments is apparent. However, it is not the only technique available at XFEL facilities. Other methods for probing plasmas can complement imaging: small-angle x-ray scattering can be used to study kinetic instabilities and blast waves with nanometer scale resolution \cite{kluge_2017, kluge_2018, gaus_2021, ordyna_2024}, grazing incidence small-angle x-ray scattering provides information on surface structures \cite{randolph_2022, randolph_2024, randolph_2025}, x-ray Thomson scattering accesses temperature and collective phenomena \cite{glenzer_2009, redmer_2023,gawne_2024, dornheim_2022}, diffraction is a standard diagnostic to study phase transitions and material deformation under high-strain \cite{McBride_2019, pandolfi_2022, pereira_2025}, among others.

\begin{figure}[!h]
\includegraphics[width=\textwidth]{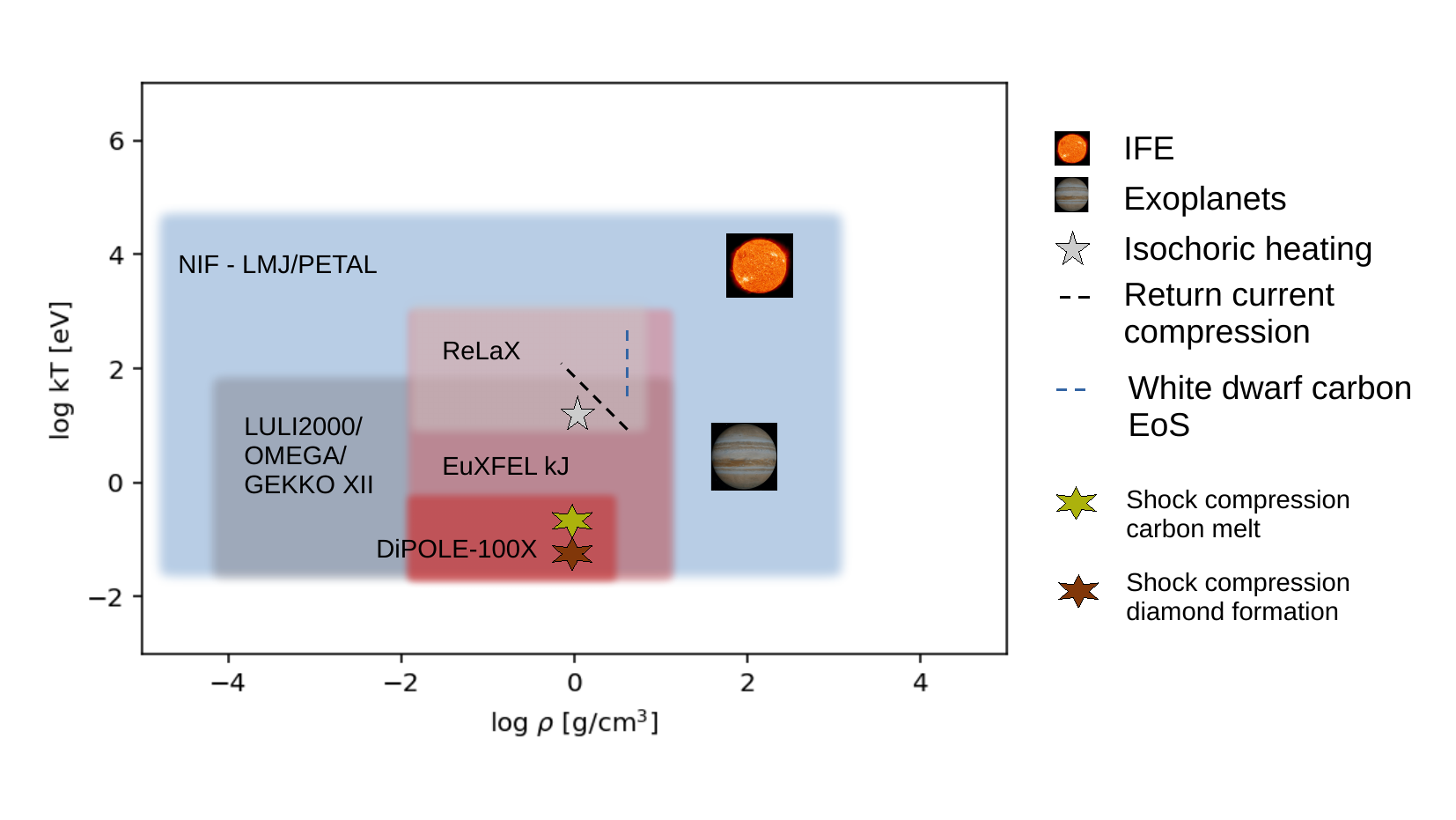}
\caption{Illustration of the phase map for carbon and hydrocarbon materials covered by different laser facilities. IFE shows the implosion point achieved at NIF in 2022 \cite{shawareb_2024}. Explanets (cite). Isochoric heating of plastic materials \cite{ohshima_2010}. Return current driven cylindrical compression \cite{lasogarcia_2024}. Carbon equation-of-state measurement at white dwarf conditions \cite{kritcher_2020}. Carbon melting and phase-coexistence \cite{kraus_2025}. Diamond formation from hydrocarbons \cite{kraus_2017}.}
\label{figure:fig_laser_phase_map}
\end{figure}

Finally, it is of interest to discuss the potential of XFELs when coupled to higher-intensity and higher-energy lasers. A workshop designed to identify the science that can be addressed at the European XFEL took place in 2024 \cite{zastrau_2024} with a follow-up in 2025. Within the German Fusion Action Plan as part of the High Tech Agenda Germany, the possibility of installing a kJ-class laser and a petawatt short-pulse laser at the European XFEL is under consideration. 

To provide an overview as to what science cases could be achieved with such a facility, we compare the conditions achievable at HED-HiBEF with other lasers used for IFE research in Figure \ref{figure:fig_laser_phase_map}. There the conditions achieved for an indirect drive ignition are represented by the "IFE" point \cite{shawareb_2024}, only reachable by Mega-Joule lasers. Other experiments at NIF studied the carbon equation of state in conditions relevant to white dwarf envelopes \cite{kritcher_2020}. While temperatures of keV are achievable with short pulse lasers, the relevant compression factor is not. Cylindrical compression of wires has shown to reach a compression of x10 in copper, and is expected to reach similar compression for hydrocarbons, the temperature at the convergence point is of 10s eV, lower than the one achieved at NIF. Isochoric heating studies of plastics in reduced mass targets was investigated with Gekko MII \cite{ohshima_2010}, at intensities $10^{17} - 10^{18}$ W cm$^{-2}$, below the maximum achievable intensity at ReLaX. In planar shocks, the study of liquid carbon was done at HED-HiBEF with DiPOLE-100X \cite{kraus_2025}, while the original diamond formation in shock compressed hydrocarbons was performed at the MEC station \cite{kraus_2017}.

A kJ-class laser would increase the pressure range currently achievable with DiPOLE-100X from $\sim$1.5 Mbar, to 10-20 Mbar, and the combination with the XFEL beam would open the path to exploring such states with unprecedented spatial and temporal resolution. At the same time, a short-pulse high-intensity petawatt laser is also considered. This would open the possibility to full equation of state measurements via a combination of imaging and x-ray Thomson scattering, extending existing equation of state measurements, for example in the case of plastic foams \cite{aglitskiy_2025}, or providing complementary precision measurements for HDC ablators \cite{millot_2018, millot_2020} beyond the current limitations of VISAR techniques. As shown in section \ref{sec:resonant_imaging}, the XFEL beam energy can be tuned to K-shell to L-shell transitions. Resonant imaging would track the evolution of ablators and material mixing with high-spatial and temporal resolution as well as temperature resolved measurements (by shifting the xfel energy across different resonances) by means of dopants in the target acting as tracers \cite{cohen_2004}. The imaging of instabilities has already been mentioned; however, a combination of small-angle x-ray scattering  together with imaging, would allow the measurement of instabilities and shocks (and shock front gradient) with spatial scales from nm to \SI{}{\um} \cite{kluge_2025}.

\section{Conclusions}
In this paper, we have introduced the hard x-ray imaging platform developed at HED-HiBEF. We have shown the setups combining it together with the ReLaX and DiPOLE-100X drivers. The platform has a spatial resolution better than 500 nm and a temporal resolution better than 50 fs. We have compared the performance of magnified x-ray imaging and Talbot imaging, showing the advantages of the latter while investigating blast waves and cylindrical compression in aluminium wires with the ReLaX laser. Tuning the XFEL energy to a resonant energy with a specific electronic transition, an increase the opacity induced by the transition of electrons from the K-shell to the L-shell has been observed, providing a diagnostic of the plasma state with specific conditions. An application to imaging of planar shocks with DiPOLE-100X has been discussed, showcasing the spatial sensitivity to probe intricacies of shock front generation and propagation. Finally, we have discussed the complementarity of this technique with other x-ray methods available at HED-HiBEF and have shown some of the science cases related to fusion that could be addressed with a kJ-class laser coupled to the XFEL beam.

\section*{Acknowledgments}
We thank the European XFEL in Schenefeld, Germany, and the HiBEF user consortium for the provision of X-ray laser time at the HED-HiBEF scientific instrument under proposal numbers 5689, as part of the HiBEF priority access, 4597 and 9796, and thank their staff for their support and the equipment provided to make this experiment possible. We thank Dr. Steve Gales from AWE for the provision of the polystyrene rod targets.
This work was partially supported by the Center for Advanced Systems Understanding (CASUS), financed by Germany’s Federal Ministry of Education and Research (BMBF) and the Saxon state government out of the State budget approved by the Saxon State Parliament. This work has received funding from the European Union's Just Transition Fund (JTF) within the project \textit{R\"ontgenlaser-Optimierung der Laserfusion} (ROLF), contract number 5086999001, co-financed by the Saxon state government out of the State budget approved by the Saxon State Parliament. Tobias Dornheim gratefully acknowledges funding from the Deutsche Forschungsgemeinschaft (DFG) via project DO 2670/1-1. The work of G. P.-C. has been supported by Research Grant No. PID2022-137632OB-I00 from the Spanish Ministry of Science and Innovation.

\section*{Roles}
A. L. G., M. M., V. B., T. T., B. N. designed the experiment. A. L. G., G. P.-C. and M.-L. H. analysed the data, A. L. G., M. M., V. B., P. V. G., M.-L. H., H. H., M. J., J. L., J.-P. N., Q. N., C. B. Q., S. V. R., L. R., A. R.-F., C. S., J. S., M. T., P. V., T. T., K. A., C. B., E. B., M. C., D. D., N. F., T. G., S. G., J. H., L. H., O. S. H., T. K., H. J. L., M. M., B. N., M. N., A. P., T. R. P., M. R., M. S., M. V. and L. Y. performed the experiments. T. T. and A. L. G. supervised the project. All authors reviewed the manuscript.

\section*{Data availability}
Data recorded for the experiment at the European XFEL are available at doi:10.22003/XFEL.EU-DATA-005689-00, doi:10.22003/XFEL.EU-DATA-004597-00, and doi:10.22003/XFEL.EU-DATA-009796-00.

\bibliography{references}

@article{Mason, title={Development of a 100 J, 10 Hz laser for compression experiments at the High Energy Density instrument at the European XFEL}, volume={6}, DOI={10.1017/hpl.2018.56}, journal={High Power Laser Science and Engineering}, author={Mason, Paul and Banerjee, Saumyabrata and Smith, Jodie and Butcher, Thomas and Phillips, Jonathan and Höppner, Hauke and Möller, Dominik and Ertel, Klaus and De Vido, Mariastefania and Hollingham, Ian and et al.}, year={2018}, pages={e65}}

@article{landen_2001,
    author = {Landen, O. L. and Farley, D. R. and Glendinning, S. G. and Logory, L. M. and Bell, P. M. and Koch, J. A. and Lee, F. D. and Bradley, D. K. and Kalantar, D. H. and Back, C. A. and Turner, R. E.},
    title = {X-ray backlighting for the National Ignition Facility (invited)},
    journal = {Review of Scientific Instruments},
    volume = {72},
    number = {1},
    pages = {627-634},
    year = {2001},
    month = {01},
    issn = {0034-6748},
    doi = {10.1063/1.1315641},
    url = {https://doi.org/10.1063/1.1315641},
    eprint = {https://pubs.aip.org/aip/rsi/article-pdf/72/1/627/19121727/627_1_online.pdf},
}

@article{barrios_2014,
    author = {Barrios, M. A. and Regan, S. P. and Fournier, K. B. and Epstein, R. and Smith, R. and Lazicki, A. and Rygg, R. and Fratanduono, D. E. and Eggert, J. and Park, H.-S. and Huntington, C. and Bradley, D. K. and Landen, O. L. and Collins, G. W.},
    title = {X-ray area backlighter development at the National Ignition Facility (invited)a)},
    journal = {Review of Scientific Instruments},
    volume = {85},
    number = {11},
    pages = {11D502},
    year = {2014},
    month = {08},
    issn = {0034-6748},
    doi = {10.1063/1.4891713},
    url = {https://doi.org/10.1063/1.4891713},
    eprint = {https://pubs.aip.org/aip/rsi/article-pdf/doi/10.1063/1.4891713/13418527/11d502_1_online.pdf},
}

@article{turk_2010,
    author = {Turk, G. and Reverdin, C. and Gontier, D. and Darbon, S. and Dujardin, C. and Ledoux, G. and Hamel, M. and Simic, V. and Normand, S.},
    title = {Development of an x-ray imaging system for the Laser Megajoule (LMJ)a)},
    journal = {Review of Scientific Instruments},
    volume = {81},
    number = {10},
    pages = {10E509},
    year = {2010},
    month = {10},
    issn = {0034-6748},
    doi = {10.1063/1.3475788},
    url = {https://doi.org/10.1063/1.3475788},
    eprint = {https://pubs.aip.org/aip/rsi/article-pdf/doi/10.1063/1.3475788/16038714/10e509_1_online.pdf},
}

@article{marshall_2021,
    author = {Marshall, F. J. and Ivancic, S. T. and Mileham, C. and Nilson, P. M. and Ruby, J. J. and Stoeckl, C. and Scheiner, B. S. and Schmitt, M. J.},
    title = {High-resolution x-ray radiography with Fresnel zone plates on the University of Rochester’s OMEGA Laser Systems},
    journal = {Review of Scientific Instruments},
    volume = {92},
    number = {3},
    pages = {033701},
    year = {2021},
    month = {03},
    issn = {0034-6748},
    doi = {10.1063/5.0034903},
    url = {https://doi.org/10.1063/5.0034903},
    eprint = {https://pubs.aip.org/aip/rsi/article-pdf/doi/10.1063/5.0034903/13862378/033701_1_online.pdf},
}

@article{bouffetier_2024,
author = {V. Bouffetier and G. P\'{e}rez-Callejo and D. Stutman and C. Stoeckl and I. A. Begishev and W. Theobald and T. Filkins and C. Mileham and L. Ceurvorst and S. R. Klein and T. Goudal and A. Casner and M. P. Valdivia},
journal = {Opt. Express},
number = {20},
pages = {34694--34709},
publisher = {Optica Publishing Group},
title = {Referenceless, grating-based, single shot X-ray phase contrast imaging with optimized laser-driven K-\&\#x03B1; sources},
volume = {32},
month = {Sep},
year = {2024},
url = {https://opg.optica.org/oe/abstract.cfm?URI=oe-32-20-34694},
doi = {10.1364/OE.516695},
}

@article{tommasini_2015,
    author = {Tommasini, R. and Field, J. E. and Hammel, B. A. and Landen, O. L. and Haan, S. W. and Aracne-Ruddle, C. and Benedetti, L. R. and Bradley, D. K. and Callahan, D. A. and Dewald, E. L. and Doeppner, T. and Edwards, M. J. and Hurricane, O. A. and Izumi, N. and Jones, O. A. and Ma, T. and Meezan, N. B. and Nagel, S. R. and Rygg, J. R. and Segraves, K. S. and Stadermann, M. and Strauser, R. J. and Town, R. P. J.},
    title = {Tent-induced perturbations on areal density of implosions at the National Ignition Facilitya)},
    journal = {Physics of Plasmas},
    volume = {22},
    number = {5},
    pages = {056315},
    year = {2015},
    month = {05},
    issn = {1070-664X},
    doi = {10.1063/1.4921218},
    url = {https://doi.org/10.1063/1.4921218},
    eprint = {https://pubs.aip.org/aip/pop/article-pdf/doi/10.1063/1.4921218/13788027/056315_1_online.pdf},
}

@article{rigon_luli_2021,
  title = {Exploring the Atwood-number dependence of the highly nonlinear Rayleigh-Taylor instability regime in high-energy-density conditions},
  author = {Rigon, G. and Albertazzi, B. and Mabey, P. and Michel, Th. and Falize, E. and Bouffetier, V. and Ceurvorst, L. and Masse, L. and Koenig, M. and Casner, A.},
  journal = {Phys. Rev. E},
  volume = {104},
  issue = {4},
  pages = {045213},
  numpages = {10},
  year = {2021},
  month = {Oct},
  publisher = {American Physical Society},
  doi = {10.1103/PhysRevE.104.045213},
  url = {https://link.aps.org/doi/10.1103/PhysRevE.104.045213}
}

@article{antonelli_2024,
    author = {Antonelli, L. and Theobald, W. and Barbato, F. and Atzeni, S. and Batani, D. and Betti, R. and Bouffetier, V. and Casner, A. and Ceurvorst, L. and Cao, D. and Ruby, J. J. and Glize, K. and Goudal, T. and Kar, A. and Khan, M. and Dearling, A. and Koenig, M. and Nilson, P. M. and Scott, R. H. H. and Turianska, O. and Wei, M. and Woolsey, N. C.},
    title = {X-ray phase-contrast imaging of strong shocks on OMEGA EP},
    journal = {Review of Scientific Instruments},
    volume = {95},
    number = {11},
    pages = {113504},
    year = {2024},
    month = {11},
    issn = {0034-6748},
    doi = {10.1063/5.0168059},
    url = {https://doi.org/10.1063/5.0168059},
    eprint = {https://pubs.aip.org/aip/rsi/article-pdf/doi/10.1063/5.0168059/20242066/113504_1_5.0168059.pdf},
}

@article{antonelli_2017,
  title = {Laser-driven shock waves studied by x-ray radiography},
  author = {Antonelli, L. and Atzeni, S. and Schiavi, A. and Baton, S. D. and Brambrink, E. and Koenig, M. and Rousseaux, C. and Richetta, M. and Batani, D. and Forestier-Colleoni, P. and Le Bel, E. and Maheut, Y. and Nguyen-Bui, T. and Ribeyre, X. and Trela, J.},
  journal = {Phys. Rev. E},
  volume = {95},
  issue = {6},
  pages = {063205},
  numpages = {8},
  year = {2017},
  month = {Jun},
  publisher = {American Physical Society},
  doi = {10.1103/PhysRevE.95.063205},
  url = {https://link.aps.org/doi/10.1103/PhysRevE.95.063205}
}

@article{wegert_2024,
    author = {Wegert, Leonard and Schreiner, Stephan and Rauch, Constantin and Albertazzi, Bruno and Bleuel, Paulina and Fröjdh, Eric and Koenig, Michel and Ludwig, Veronika and Martynenko, Artem S. and Meyer, Pascal and Mozzanica, Aldo and Müller, Michael and Neumayer, Paul and Schneider, Markus and Triantafyllidis, Angelos and Zielbauer, Bernhard and Anton, Gisela and Michel, Thilo and Funk, Stefan},
    title = {Demonstrating grating-based phase-contrast imaging of laser-driven shock waves},
    journal = {Matter and Radiation at Extremes},
    volume = {9},
    number = {4},
    pages = {047803},
    year = {2024},
    month = {06},
    issn = {2468-2047},
    doi = {10.1063/5.0200440},
    url = {https://doi.org/10.1063/5.0200440},
    eprint = {https://pubs.aip.org/aip/mre/article-pdf/doi/10.1063/5.0200440/20004493/047803_1_5.0200440.pdf},
}

@article{wood_2018,
author={Wood, J. C. and Chapman, D. J. and Poder, K. and Lopes, N. C. and Rutherford, M. E. and White, T. G. and Albert, F. and Behm, K. T. and Booth, N. and Bryant, J. S. J. and Foster, P. S. and Glenzer, S. and Hill, E. and Krushelnick, K. and Najmudin, Z. and Pollock, B. B. and Rose, S. and Schumaker, W. and Scott, R. H. H. and Sherlock, M. and Thomas, A. G. R. and Zhao, Z. and Eakins, D. E. and Mangles, S. P. D.},
title={Ultrafast Imaging of Laser Driven Shock Waves using Betatron X-rays from a Laser Wakefield Accelerator},
journal={Scientific Reports},
year={2018},
month={Jul},
day={20},
volume={8},
number={1},
pages={11010},
issn={2045-2322},
doi={10.1038/s41598-018-29347-0},
url={https://doi.org/10.1038/s41598-018-29347-0}
}

@article{do_2021,
    author = {Do, A. and Angulo, A. M. and Hall, G. N. and Nagel, S. R. and Izumi, N. and Kozioziemski, B. J. and McCarville, T. and Ayers, J. M. and Bradley, D. K.},
    title = {X-ray imaging of Rayleigh–Taylor instabilities using Fresnel zone plate at the National Ignition Facility},
    journal = {Review of Scientific Instruments},
    volume = {92},
    number = {5},
    pages = {053511},
    year = {2021},
    month = {05},
    issn = {0034-6748},
    doi = {10.1063/5.0043682},
    url = {https://doi.org/10.1063/5.0043682},
    eprint = {https://pubs.aip.org/aip/rsi/article-pdf/doi/10.1063/5.0043682/14125690/053511_1_online.pdf},
}

@article{galtier_2025,
author={Galtier, Eric and Lee, Hae Ja and Khaghani, Dimitri and Boiadjieva, Nina and McGehee, Peregrine and Arnott, Ariel and Arnold, Brice and Berboucha, Meriame and Cunningham, Eric and Czapla, Nick and Dyer, Gilliss and Ettelbrick, Robert and Hart, Philip and Heimann, Philip and Welch, Marc and Makita, Mikako and Gleason, Arianna E. and Pandolfi, Silvia and Sakdinawat, Anne and Liu, Yanwei and Wojcik, Michael J. and Hodge, Daniel
and Sandberg, Richard and Valdivia, Maria Pia and Bouffetier, Victorien and P{\'e}rez-Callejo, Gabriel and Seiboth, Frank and Nagler, Bob},
title={X-ray microscopy and talbot imaging with the matter in extreme conditions X-ray imager at LCLS},
journal={Scientific Reports},
year={2025},
month={Mar},
day={04},
volume={15},
number={1},
pages={7588},
issn={2045-2322},
doi={10.1038/s41598-025-91989-8},
url={https://doi.org/10.1038/s41598-025-91989-8}
}

@article{koliyadu_2025,
author = "Koliyadu, Jayanath C. P. and Mo{\v{s}}ko, Daniel and Asimakopoulou, Eleni Myrto and Bellucci, Valerio and Birn{\v{s}}teinov{\'{a}}, {\v{S}}arlota and Bean, Richard and Letrun, Romain and Kim, Chan and Kirkwood, Henry and Giovanetti, Gabriele and Jardon, Nerea and Szuba, Janusz and Guest, Trey and Koch, Andreas and Gr{\"{u}}nert, Jan and Szeles, Peter and Villanueva-Perez, Pablo and Reuter, Fabian and Ohl, Claus-Dieter and Noack, Mike Andreas and Garcia-Moreno, Francisco and Kuglerov{\'{a}}-Valdov{\'{a}}, Zuzana and Juha, Libor and Nikl, Martin and Yashiro, Wataru and Soyama, Hitoshi and Eakins, Daniel and Korsunsky, Alexander M. and Uli{\v{c}}n{\'{y}}, Jozef and Meents, Alke and Chapman, Henry N. and Mancuso, Adrian P. and Sato, Tokushi and Vagovi{\v{c}}, Patrik",
title = "{Development of MHz X-ray phase contrast imaging at the European XFEL}",
journal = "Journal of Synchrotron Radiation",
year = "2025",
volume = "32",
number = "1",
pages = "17--28",
month = "Jan",
doi = {10.1107/S160057752400986X},
url = {https://doi.org/10.1107/S160057752400986X},
}

@article{tschentscher_2017,
AUTHOR = {Tschentscher, Thomas and Bressler, Christian and Grünert, Jan and Madsen, Anders and Mancuso, Adrian P. and Meyer, Michael and Scherz, Andreas and Sinn, Harald and Zastrau, Ulf},
TITLE = {Photon Beam Transport and Scientific Instruments at the European XFEL},
JOURNAL = {Applied Sciences},
VOLUME = {7},
YEAR = {2017},
NUMBER = {6},
ARTICLE-NUMBER = {592},
URL = {https://www.mdpi.com/2076-3417/7/6/592},
ISSN = {2076-3417},
DOI = {10.3390/app7060592}
}

@phdthesis{bouffetier_thesis_2021,
  TITLE = {{D{\'e}veloppement de l'interf{\'e}rom{\'e}trie X et application {\`a} l'imagerie par contraste de phase de plasmas denses et turbulents}},
  AUTHOR = {Bouffetier, Victorien},
  URL = {https://theses.hal.science/tel-03634981},
  NUMBER = {2021BORD0292},
  SCHOOL = {{Universit{\'e} de Bordeaux}},
  YEAR = {2021},
  MONTH = Dec,
  TYPE = {Theses},
  HAL_ID = {tel-03634981},
  HAL_VERSION = {v1},
}

@article{parisuana_2025,
    author = {Parisuaña, C. and Valdivia, M. P. and Bouffetier, V. and Kurzer-Ogul, K. and Pérez-Callejo, G. and Bott-Suzuki, S. and Casner, A. and Christiansen, N. S. and Czapla, N. and Eder, D. and Galtier, E. and Glenzer, S. H. and Goudal, T. and Haines, B. M. and Hodge, D. and Ikeya, M. and Izquierdo, L. and Khaghani, D. and Kim, Y. and Klein, S. and Koniges, A. and Lee, H. J. and Leininger, M. and Leong, A. F. T. and Lester, R. S. and Makita, M. and Mancelli, D. and Martin, W. M. and Nagler, B. and Sandberg, R. L. and Truong, A. and Vescovi, M. and Gleason, A. E. and Kozlowski, P. M.},
    title = {Shock propagation in aerogel and TPP foams for inertial fusion energy target design},
    journal = {Physics of Plasmas},
    volume = {32},
    number = {8},
    pages = {082707},
    year = {2025},
    month = {08},
    issn = {1070-664X},
    doi = {10.1063/5.0273572},
    url = {https://doi.org/10.1063/5.0273572},
    eprint = {https://pubs.aip.org/aip/pop/article-pdf/doi/10.1063/5.0273572/20670858/082707_1_5.0273572.pdf},
}

@article{hodge_2025,
    author = {Hodge, D. S. and Leong, A. F. T. and Kurzer-Ogul, K. and Pandolfi, S. and Montgomery, D. S. and Shang, J. and Aluie, H. and Marchesini, S. and Liu, Y. and Li, K. and Sakdinawat, A. and Galtier, E. C. and Nagler, B. and Lee, H. J. and Cunningham, E. F. and Carver, T. E. and Bolme, C. A. and Ramos, K. J. and Khaghani, D. and Kozlowski, P. M. and Gleason, A. E. and Sandberg, R. L.},
    title = {Single-shot in-line x-ray phase-contrast imaging of void-shockwave interactions in fusion energy materials},
    journal = {Physics of Plasmas},
    volume = {32},
    number = {8},
    pages = {083903},
    year = {2025},
    month = {08},
    issn = {1070-664X},
    doi = {10.1063/5.0272820},
    url = {https://doi.org/10.1063/5.0272820},
    eprint = {https://pubs.aip.org/aip/pop/article-pdf/doi/10.1063/5.0272820/20663945/083903_1_5.0272820.pdf},
}

@article{schropp_2015,
author={Schropp, Andreas and Hoppe, Robert and Meier, Vivienne and Patommel, Jens and Seiboth, Frank and Ping, Yuan and Hicks, Damien G. and Beckwith, Martha A. and Collins, Gilbert W. and Higginbotham, Andrew and Wark, Justin S. and Lee, Hae Ja and Nagler, Bob and Galtier, Eric C. and Arnold, Brice and Zastrau, Ulf and Hastings, Jerome B. and Schroer, Christian G.},
title={Imaging Shock Waves in Diamond with Both High Temporal and Spatial Resolution at an XFEL},
journal={Scientific Reports},
year={2015},
month={Jun},
day={18},
volume={5},
number={1},
pages={11089},
issn={2045-2322},
doi={10.1038/srep11089},
url={https://doi.org/10.1038/srep11089}
}

@article{nagler_2016,
    author = {Nagler, Bob and Schropp, Andreas and Galtier, Eric C. and Arnold, Brice and Brown, Shaughnessy B. and Fry, Alan and Gleason, Arianna and Granados, Eduardo and Hashim, Akel and Hastings, Jerome B. and Samberg, Dirk and Seiboth, Frank and Tavella, Franz and Xing, Zhou and Lee, Hae Ja and Schroer, Christian G.},
    title = {The phase-contrast imaging instrument at the matter in extreme conditions endstation at LCLS},
    journal = {Review of Scientific Instruments},
    volume = {87},
    number = {10},
    pages = {103701},
    year = {2016},
    month = {10},
    issn = {0034-6748},
    doi = {10.1063/1.4963906},
    url = {https://doi.org/10.1063/1.4963906},
    eprint = {https://pubs.aip.org/aip/rsi/article-pdf/doi/10.1063/1.4963906/14056366/103701_1_online.pdf},
}

@article{hodge_2022,
author = {Daniel S. Hodge and Andrew F. T. Leong and Silvia Pandolfi and Kelin Kurzer-Ogul and David S. Montgomery and Hussein Aluie and Cindy Bolme and Thomas Carver and Eric Cunningham and Chandra B. Curry and Matthew Dayton and Franz-Joseph Decker and Eric Galtier and Philip Hart and Dimitri Khaghani and Hae Ja Lee and Kenan Li and Yanwei Liu and Kyle Ramos and Jessica Shang and Sharon Vetter and Bob Nagler and Richard L. Sandberg and Arianna E. Gleason},
journal = {Opt. Express},
number = {21},
pages = {38405--38422},
publisher = {Optica Publishing Group},
title = {Multi-frame, ultrafast, x-ray microscope for imaging shockwave dynamics},
volume = {30},
month = {Oct},
year = {2022},
url = {https://opg.optica.org/oe/abstract.cfm?URI=oe-30-21-38405},
doi = {10.1364/OE.472275},
}

@article{tanaka_2025,
    author = {Tanaka, D. and Sawada, H. and Idesaka, T. and Nakatsuji, C. and Matsuura, S. and Sato, T. and Somekawa, T. and Yabuuchi, T. and Miyanishi, K. and Sueda, K. and Inubushi, Y. and Sentoku, Y. and Shimizu, T. and Shingubara, S. and Kawasaki, K. and Ozaki, N. and Yamanoi, K. and Shigemori, K.},
    title = {Experimental investigation on nanowire array irradiated with ultrahigh intensity laser at x-ray free electron laser facility SACLA: Fabrication of nanowire array target and its application to ultrafast time-resolved measurements},
    journal = {Journal of Applied Physics},
    volume = {137},
    number = {12},
    pages = {125901},
    year = {2025},
    month = {03},
    issn = {0021-8979},
    doi = {10.1063/5.0251649},
    url = {https://doi.org/10.1063/5.0251649},
    eprint = {https://pubs.aip.org/aip/jap/article-pdf/doi/10.1063/5.0251649/20457197/125901_1_5.0251649.pdf},
}

@article{makarov_2023,
    author = {Makarov, Sergey and Dyachkov, Sergey and Pikuz, Tatiana and Katagiri, Kento and Nakamura, Hirotaka and Zhakhovsky, Vasily and Inogamov, Nail and Khokhlov, Victor and Martynenko, Artem and Albertazzi, Bruno and Rigon, Gabriel and Mabey, Paul and Hartley, Nicholas J. and Inubushi, Yuichi and Miyanishi, Kohei and Sueda, Keiichi and Togashi, Tadashi and Yabashi, Makina and Yabuuchi, Toshinori and Okuchi, Takuo and Kodama, Ryosuke and Pikuz, Sergey and Koenig, Michel and Ozaki, Norimasa},
    title = {Direct imaging of shock wave splitting in diamond at Mbar pressure},
    journal = {Matter and Radiation at Extremes},
    volume = {8},
    number = {6},
    pages = {066601},
    year = {2023},
    month = {09},
    issn = {2468-2047},
    doi = {10.1063/5.0156681},
    url = {https://doi.org/10.1063/5.0156681},
    eprint = {https://pubs.aip.org/aip/mre/article-pdf/doi/10.1063/5.0156681/18123824/066601_1_5.0156681.pdf},
}

@article{lasogarcia_2024,
author={Laso Garcia, Alejandro and Yang, Long and Bouffetier, Victorien and Appel, Karen and Baehtz, Carsten and Hagemann, Johannes and H{\"o}ppner, Hauke and Humphries, Oliver and Kluge, Thomas and Mishchenko, Mikhail and Nakatsutsumi, Motoaki and Pelka, Alexander and Preston, Thomas R. and Randolph, Lisa and Zastrau, Ulf and Cowan, Thomas E. and Huang, Lingen and Toncian, Toma},
title={Cylindrical compression of thin wires by irradiation with a Joule-class short-pulse laser},
journal={Nature Communications},
year={2024},
month={Sep},
day={12},
volume={15},
number={1},
pages={7896},
issn={2041-1723},
doi={10.1038/s41467-024-52232-6},
url={https://doi.org/10.1038/s41467-024-52232-6}
}

@article{zastrau_2021,
author = "Zastrau, Ulf and Appel, Karen and Baehtz, Carsten and Baehr, Oliver and Batchelor, Lewis and Bergh{\"{a}}user, Andreas and Banjafar, Mohammadreza and Brambrink, Erik and Cerantola, Valerio and Cowan, Thomas E. and Damker, Horst and Dietrich, Steffen and Di Dio Cafiso, Samuele and Dreyer, J{\"{o}}rn and Engel, Hans-Olaf and Feldmann, Thomas and Findeisen, Stefan and Foese, Manon and Fulla-Marsa, Daniel and G{\"{o}}de, Sebastian and Hassan, Mohammed and Hauser, Jens and Herrmannsd{\"{o}}rfer, Thomas and H{\"{o}}ppner, Hauke and Kaa, Johannes and Kaever, Peter and Kn{\"{o}}fel, Klaus and Kon{\^{o}}pkov{\'{a}}, Zuzana and Laso Garc{\'\i}a, Alejandro and Liermann, Hanns-Peter and Mainberger, Jona and Makita, Mikako and Martens, Eike-Christian and McBride, Emma E. and M{\"{o}}ller, Dominik and Nakatsutsumi, Motoaki and Pelka, Alexander and Plueckthun, Christian and Prescher, Clemens and Preston, Thomas R. and R{\"{o}}per, Michael and Schmidt, Andreas and Seidel, Wolfgang and Schwinkendorf, Jan-Patrick and Schoelmerich, Markus O. and Schramm, Ulrich and Schropp, Andreas and Strohm, Cornelius and Sukharnikov, Konstantin and Talkovski, Peter and Thorpe, Ian and Toncian, Monika and Toncian, Toma and Wollenweber, Lennart and Yamamoto, Shingo and Tschentscher, Thomas",
title = "{The High Energy Density Scientific Instrument at the European XFEL}",
journal = "Journal of Synchrotron Radiation",
year = "2021",
volume = "28",
number = "5",
pages = "1393--1416",
month = "Sep",
doi = {10.1107/S1600577521007335},
url = {https://doi.org/10.1107/S1600577521007335},
}

@article{lasogarcia_2021,
author={Laso Garcia, A. and H{\"o}ppner, H. and Pelka, A. and B{\"a}htz, C. and Brambrink, E. and Di Dio Cafiso, S. and Dreyer, J. and G{\"o}de, S. and Hassan, M. and Kluge, T. and Liu, J. and Makita, M. and M{\"o}ller, D. and Nakatsutsumi, M. and Preston, T. R. and Priebe, G. and Schlenvoigt, H.-P. and Schwinkendorf, J.-P. and {\v{S}}m{\'i}d, M. and Talposi, A.-M. and Toncian, M. and Zastrau, U. and Schramm, U. and Cowan, T. E. and Toncian, T.},
title={ReLaX: the Helmholtz International Beamline for Extreme Fields high-intensity short-pulse laser driver for relativistic laser--matter interaction and strong-field science using the high energy density instrument at the European X-ray free electron laser facility},
journal={High Power Laser Science and Engineering},
year={2021},
edition={2021/10/18},
publisher={Cambridge University Press},
volume={9},
pages={e59},
note={e59},
issn={2095-4719},
doi={10.1017/hpl.2021.47},
url={https://www.cambridge.org/core/product/2E9CD2B2288414E5785888C2354B3E7E},
url={https://doi.org/10.1017/hpl.2021.47}
}

@ARTICLE{teague_1982,
	author = {TEAGUE, MICHAEL REED},
    title = {Irradiance moments: their propagation and use for unique retrieval of phase},
	year = {1982},
	journal = {J OPT SOC AM},
	volume = {V 72},
	number = {N 9},
	pages = {1199 – 1209},
	doi = {10.1364/josa.72.001199},
	url = {https://www.scopus.com/inward/record.uri?eid=2-s2.0-0020178653&doi=10.1364%2fjosa.72.001199&partnerID=40&md5=b5ef1c182341197ffbaf7cc95e9e168a},
	type = {Article},
	publication_stage = {Final},
	source = {Scopus},
	note = {Cited by: 237}
}

@article{paganin_2002,
author = {Paganin, D. and Mayo, S. C. and Gureyev, T. E. and Miller, P. R. and Wilkins, S. W.},
title = {Simultaneous phase and amplitude extraction from a single defocused image of a homogeneous object},
journal = {Journal of Microscopy},
volume = {206},
number = {1},
pages = {33-40},
doi = {https://doi.org/10.1046/j.1365-2818.2002.01010.x},
url = {https://onlinelibrary.wiley.com/doi/abs/10.1046/j.1365-2818.2002.01010.x},
eprint = {https://onlinelibrary.wiley.com/doi/pdf/10.1046/j.1365-2818.2002.01010.x},
year = {2002}
}

@article{paganin_2020,
doi = {10.1088/2040-8986/abbab9},
url = {https://doi.org/10.1088/2040-8986/abbab9},
year = {2020},
month = {oct},
publisher = {IOP Publishing},
volume = {22},
number = {11},
pages = {115607},
author = {Paganin, David M and Favre-Nicolin, Vincent and Mirone, Alessandro and Rack, Alexander and Villanova, Julie and Olbinado, Margie P and Fernandez, Vincent and da Silva, Julio C and Pelliccia, Daniele},
title = {Boosting spatial resolution by incorporating periodic boundary conditions into single-distance hard-x-ray phase retrieval},
journal = {Journal of Optics}
}

@article{huhn_2022,
author = {Simon Huhn and Leon Merten Lohse and Jens Lucht and Tim Salditt},
journal = {Opt. Express},
number = {18},
pages = {32871--32886},
publisher = {Optica Publishing Group},
title = {Fast algorithms for nonlinear and constrained phase retrieval in near-field X-ray holography based on Tikhonov regularization},
volume = {30},
month = {Aug},
year = {2022},
url = {https://opg.optica.org/oe/abstract.cfm?URI=oe-30-18-32871},
doi = {10.1364/OE.462368},
}

@article{valdivia_2018,
author = {Maria Pia Valdivia and Dan Stutman and Christian Stoeckl and Chad Mileham and Ildar A. Begishev and Jake Bromage and Sean P. Regan},
journal = {Appl. Opt.},
number = {2},
pages = {138--145},
publisher = {Optica Publishing Group},
title = {Talbot\&\#x2013;Lau x-ray deflectometry phase-retrieval methods for electron density diagnostics in high-energy density experiments},
volume = {57},
month = {Jan},
year = {2018},
url = {https://opg.optica.org/ao/abstract.cfm?URI=ao-57-2-138},
doi = {10.1364/AO.57.000138},
}

@article{bouffetier_2020,
author = {V. Bouffetier and L. Ceurvorst and M. P. Valdivia and F. Dorchies and S. Hulin and T. Goudal and D. Stutman and A. Casner},
journal = {Appl. Opt.},
number = {27},
pages = {8380--8387},
publisher = {Optica Publishing Group},
title = {Proof-of-concept Talbot--Lau x-ray interferometry with a high-intensity, high-repetition-rate, laser-driven K-alpha source},
volume = {59},
month = {Sep},
year = {2020},
url = {https://opg.optica.org/ao/abstract.cfm?URI=ao-59-27-8380},
doi = {10.1364/AO.398839},
}

@article{perezcallejo_2023,
author={P{\'e}rez-Callejo, G. and Bouffetier, V. and Ceurvorst, L. and Goudal, T. and Klein, S. R. and Svyatskiy, D. and Holec, M. and Perez-Martin, P. and Falk, K. and Casner, A. and Weber, T. E. and Kagan, G. and Valdivia, M. P.},
title={Phase imaging of irradiated foils at the OMEGA EP facility using phase-stepping X-ray Talbot--Lau deflectometry},
journal={High Power Laser Science and Engineering},
year={2023},
edition={2023/05/26},
publisher={Cambridge University Press},
volume={11},
pages={e49},
note={e49},
issn={2095-4719},
doi={10.1017/hpl.2023.44},
url={https://www.cambridge.org/core/product/53AEE63F6918AF67A9FFA2C23C5377A8},
url={https://doi.org/10.1017/hpl.2023.44}
}

@article{valdivia_2021,
    author = {Valdivia, M. P. and Stutman, D. and Stoeckl, C. and Theobald, W. and Collins, G. W., IV and Bouffetier, V. and Vescovi, M. and Mileham, C. and Begishev, I. A. and Klein, S. R. and Melean, R. and Muller, S. and Zou, J. and Veloso, F. and Casner, A. and Beg, F. N. and Regan, S. P.},
    title = {Talbot-Lau x-ray deflectometer: Refraction-based HEDP imaging diagnostic},
    journal = {Review of Scientific Instruments},
    volume = {92},
    number = {6},
    pages = {065110},
    year = {2021},
    month = {06},
    issn = {0034-6748},
    doi = {10.1063/5.0043655},
    url = {https://doi.org/10.1063/5.0043655},
    eprint = {https://pubs.aip.org/aip/rsi/article-pdf/doi/10.1063/5.0043655/15929165/065110_1_online.pdf},
}

@article{perezcallejo_2022,
    author = {Pérez-Callejo, G. and Bouffetier, V. and Ceurvorst, L. and Goudal, T. and Valdivia, M. P. and Stutman, D. and Casner, A.},
    title = {TIA: A forward model and analyzer for Talbot interferometry experiments of dense plasmas},
    journal = {Physics of Plasmas},
    volume = {29},
    number = {4},
    pages = {043901},
    year = {2022},
    month = {04},
    issn = {1070-664X},
    doi = {10.1063/5.0085822},
    url = {https://doi.org/10.1063/5.0085822},
    eprint = {https://pubs.aip.org/aip/pop/article-pdf/doi/10.1063/5.0085822/16613121/043901_1_online.pdf},
}

@article{batani_2024,
author={Batani, Katarzyna Liliana and Malko, Sophia and Touati, Michael and Feugeas, Jean-Luc and Lad, Amit D. and Jana, Kamalesh and Kumar, G. Ravindra and Raffestin, Didier and Turianska, Olena and Khaghani, Dimitri and Tentori, Alessandro and Mancelli, Donaldi and Martynenko, Artem S. and Pikuz, Sergey and Benocci, Roberto and Volpe, Luca and Zeraouli, Ghassan and Perez Hernandez, Jose Antonio and Garcia, Enrique and Narayanan, Venkatakrishnan and Santos, Joao and Batani, Dimitri},
title={Characterization of blast waves induced by femtosecond laser irradiation in solid targets},
journal={High Power Laser Science and Engineering},
year={2024},
edition={2024/11/04},
publisher={Cambridge University Press},
volume={12},
pages={e59},
note={e59},
issn={2095-4719},
doi={10.1017/hpl.2024.36},
url={https://www.cambridge.org/core/product/F1D2A97D6C40C4A4DB85084D784A2A44},
url={https://doi.org/10.1017/hpl.2024.36}
}

@article{santos_2007,
    author = {Santos, J. J. and Debayle, A. and Nicolaï, Ph. and Tikhonchuk, V. and Manclossi, M. and Batani, D. and Guemnie-Tafo, A. and Faure, J. and Malka, V. and Honrubia, J. J.},
    title = {Fast-electron transport and induced heating in aluminum foils},
    journal = {Physics of Plasmas},
    volume = {14},
    number = {10},
    pages = {103107},
    year = {2007},
    month = {10},
    issn = {1070-664X},
    doi = {10.1063/1.2790893},
    url = {https://doi.org/10.1063/1.2790893},
    eprint = {https://pubs.aip.org/aip/pop/article-pdf/doi/10.1063/1.2790893/15764924/103107_1_online.pdf},
}

@article{jakubowska_2017,
doi = {10.1209/0295-5075/119/35001},
url = {https://doi.org/10.1209/0295-5075/119/35001},
year = {2017},
month = {oct},
publisher = {EDP Sciences, IOP Publishing and Società Italiana di Fisica},
volume = {119},
number = {3},
pages = {35001},
author = {Jakubowska, K. and Batani, D. and Feugeas, J.-F. and Forestier-Colleoni, P. and Hulin, S. and Nicolaï, P. and Santos, J. J. and Flacco, A. and Vauzour, B. and Malka, V.},
title = {Generation of high pressures by short-pulse low-energy laser irradiation},
journal = {Europhysics Letters},
}

@article{santos_2017,
doi = {10.1088/1367-2630/aa806b},
url = {https://doi.org/10.1088/1367-2630/aa806b},
year = {2017},
month = {oct},
publisher = {IOP Publishing},
volume = {19},
number = {10},
pages = {103005},
author = {Santos, J J and Vauzour, B and Touati, M and Gremillet, L and Feugeas, J-L and Ceccotti, T and Bouillaud, R and Deneuville, F and Floquet, V and Fourment, C and Hadj-Bachir, M and Hulin, S and Morace, A and Nicolaï, Ph and d’Oliveira, P and Reau, F and Samaké, A and Tcherbakoff, O and Tikhonchuk, V T and Veltcheva, M and Batani, D},
title = {Isochoric heating and strong blast wave formation driven by fast electrons in solid-density targets},
journal = {New Journal of Physics},
}

@article{long_2025,
    author = {Yang, L. and Herbert, M.-L. and Baehtz, C. and Bouffetier, V. and Brambrink, E. and Dornheim, T. and Fefeu, N. and Gawne, T. and Goede, S. and Hagemann, J. and Höppner, H. and Huang, L. G. and Humphries, O. and Kluge, T. and Kraus, D. and Lütgert, J. and Naedler, J.-P. and Nakatsutsumi, M. and Pelka, A. and Preston, T. R. and Qu, C. B. and Rahul, S. V. and Randolph, L. and Redmer, R. and Rehwald, M. and Santos, J. J. and Šmíd, M. and Schramm, U. and Schwinkendorf, J.-P. and Vescovi, M. and Zastrau, U. and Zeil, K. and Laso Garcia, A. and Toncian, T. and Cowan, T. E.},
    title = {Scaling of thin wire cylindrical compression with material, diameter, and laser energy after 100 fs Joule surface heating},
    journal = {Matter and Radiation at Extremes},
    volume = {11},
    number = {1},
    pages = {017604},
    year = {2025},
    month = {12},
    issn = {2468-2047},
    doi = {10.1063/5.0291405},
    url = {https://doi.org/10.1063/5.0291405},
    eprint = {https://pubs.aip.org/aip/mre/article-pdf/doi/10.1063/5.0291405/20824755/017604_1_5.0291405.pdf},
}

@article{huang_2025,
    author = {Huang, Lingen and Šmíd, Michal and Yang, Long and Humphries, Oliver and Hagemann, Johannes and Engler, Thea and Pan, Xiayun and Cui, Yangzhe and Kluge, Thomas and Aguilar, Ritz and Baehtz, Carsten and Brambrink, Erik and Eren, Engin and Falk, Katerina and Laso Garcia, Alejandro and Göde, Sebastian and Gutt, Christian and Hassan, Mohamed and Heuser, Philipp and Höppner, Hauke and Kozlova, Michaela and Lu, Wei and Metzkes-Ng, Josefine and Masruri, Masruri and Mishchenko, Mikhail and Nakatsutsumi, Motoaki and Ota, Masato and Öztürk, Özgül and Pelka, Alexander and Prencipe, Irene and Preston, Thomas R. and Randolph, Lisa and Rehwald, Martin and Schlenvoigt, Hans-Peter and Schramm, Ulrich and Schwinkendorf, Jan-Patrick and Starke, Sebastian and Štefaníková, Radka and Thiessenhusen, Erik and Toncian, Monika and Toncian, Toma and Vorberger, Jan and Zastrau, Ulf and Zeil, Karl and Cowan, Thomas E.},
    title = {Demonstration of full-scale spatiotemporal diagnostics of solid-density plasmas driven by an ultra-short relativistic laser pulse using an X-ray free-electron laser},
    journal = {Matter and Radiation at Extremes},
    volume = {11},
    number = {1},
    pages = {017201},
    year = {2025},
    month = {10},
    issn = {2468-2047},
    doi = {10.1063/5.0279974},
    url = {https://doi.org/10.1063/5.0279974},
    eprint = {https://pubs.aip.org/aip/mre/article-pdf/doi/10.1063/5.0279974/20771743/017201_1_5.0279974.pdf},
}

@Article{duchateau_2025,
author={Duchateau, Guillaume
and Pradel, Pierre
and Bourdineaud, Nicolas
and H{\'e}bert, David
and Malaise, Fr{\'e}d{\'e}ric},
title={Theoretical influence of ablator thickness on laser induced hydrodynamics in materials},
journal={Applied Physics A},
year={2025},
month={Apr},
day={10},
volume={131},
number={5},
pages={355},
issn={1432-0630},
doi={10.1007/s00339-025-08487-x},
url={https://doi.org/10.1007/s00339-025-08487-x}
}

@article{hurricane_2024,
title = {How ignition and target gain >1 were achieved in inertial fusion},
journal = {High Energy Density Physics},
volume = {53},
pages = {101157},
year = {2024},
issn = {1574-1818},
doi = {https://doi.org/10.1016/j.hedp.2024.101157},
url = {https://www.sciencedirect.com/science/article/pii/S157418182400082X},
author = {O.A. Hurricane}
}

@article{hurricane_2025,
doi = {10.1088/1361-6587/ad994f},
url = {https://doi.org/10.1088/1361-6587/ad994f},
year = {2024},
month = {dec},
publisher = {IOP Publishing},
volume = {67},
number = {1},
pages = {015019},
author = {Hurricane, O A and Allen, A and Bachmann, B L and Baker, K L and Baxamusa, S and Bhandarkar, S D and Biener, J and Bionta, S R M and Braun, T and Briggs, T and Brunton, G and Casey, D T and Chapman, T and Choate, C and Clark, D S and Dewald, E and DiNicola, J-M and Divol, L and Do, A and Fehrenbach, T and Fittinghoff, D N and Gatu Johnson, M and Geppert Kleinrath, H and Geppert Kleinrath, V and Haan, S and Hilsabeck, T J and Hinkel, D E and Hohenberger, M and Humbird, K D and Izumi, N and Kong, C and Kritcher, A L and Landen, O L and Lindl, J and MacGowan, B J and Mackinnon, A J and Maclaren, S A and Marinak, M and Meeuwsen, R and Michel, P and Milovich, J and Meaney, K and Millot, M and Moody, J D and Moore, A S and Nikroo, A and Nora, R and Pak, A and Ralph, J E and Ratledge, M and Ross, J S and Rubery, M S and Schlossberg, D J and Schmit, P F and Sepke, S M and Smalyuk, V and Spears, B K and Springer, P T and Stadermann, M and Strozzi, D J and Suratwala, T I and Tommasini, R and Town, R P J and Weber, C R and Wild, C and Van Wonterghem, B and Woodworth, B and Wu, J and Young, C V and Zylstra, A B},
title = {Present understanding of ignition and gain using indirect-drive inertial confinement fusion target designs on the U.S. National Ignition Facility},
journal = {Plasma Physics and Controlled Fusion}
}

@article{lucht_2025,
    author = "Lucht, Jens and Meyer, Paul and Lohse, Leon Merten and Salditt, Tim",
    title = "{{\it HoToPy}: a toolbox for X-ray holo-tomography in Python}",
    journal = "Journal of Synchrotron Radiation",
    year = "2025",
    volume = "32",
    number = "6",
    pages = "1586--1594",
    month = "Nov",
    doi = {10.1107/S1600577525008550},
    url = {https://doi.org/10.1107/S1600577525008550},
}

\end{document}